\documentclass[11pt]{article}
\usepackage{moriond,epsfig}
\def\Journal#1#2#3#4{{\it #1} {\bf #2}, #3 (#4)}
\def\prd#1#2#3{{\it Phys.\ Rev.} {\bf D#1}, #2 (#3)}
\def\prc#1#2#3{{\it Phys.\ Rev.} {\bf C#1}, #2 (#3)}

\def\PRD[1][2][3]{{\em Phys. Rev.} D}

\def\Dm2{\Delta m^2}
\def\nebar{\overline{\nu_e}}
\def\nmbar{\overline{\nu_\mu}}
\def\ra{\rightarrow}

\def\gsim{\;\raisebox{-.6ex}{$\stackrel{>}{\sim}$}\;}
\newcommand{\optbar}[1]{\shortstack{{\tiny (\rule[.4ex]{1em}{.1mm})} 
  \\ [-.7ex] $#1$}}		
\newcommand{\Eq}[1]{Eq.~(\ref{eq#1})}
\def\be{\begin{equation}}
\def\ee{\end{equation}}
\def\bea{\begin{eqnarray}}
\def\eea{\end{eqnarray}}

\begin{document}
\vspace*{4cm}
\title{TENSIONS WITH THE THREE-NEUTRINO PARADIGM \footnote{FERMILAB-CONF-12-384-T. Published in  {\em 2012 Electroweak Interactions and Unified Theories} (Proceedings of the Electroweak Session of the 47th Rencontres de Moriond), eds.\ E. Aug\'{e}, J. Dumarchez, J.-M. Fr\`{e}re, L. Iconomidou-Fayard, and J. Tr\^{a}n Thanh V\^{a}n (ARISF, 2012) p.191.}}

\author{B. KAYSER }
\address{Theoretical Physics Department, Fermilab, P.O. Box 500,\\
Batavia, IL 60510  USA}

\maketitle\abstracts{
We review the tensions with the standard three-neutrino framework for neutrino oscillation. These tensions hint at the possible existence of sterile neutrinos. We briefly describe some of the diverse ideas for probing the existence of such neutrinos in a definitive way through future experiments. }

\section{The Tensions}

Not all of the neutrino data are successfully described by the standard three-neutrino paradigm. This paradigm includes only three neutrino mass eigenstates, $\nu_1,\; \nu_2$, and $\nu_3$, with squared-masses separated by the splittings $|\Dm2_{21}| \equiv | m^2_2 - m^2_1 | = 7.5 \times 10^{-5}$ eV$^2$~~\cite{refA} and  $|\Dm2_{32}| \equiv | m^2_3 - m^2_2 | = 2.4 \times 10^{-3}$ eV$^2$. \cite{refB} However, there are hints, coming from a variety of sources, that nature may contain more than three neutrino mass eigenstates, and squared-mass splittings significantly larger than the measured $|\Dm2_{21}|$ and $|\Dm2_{32}|$. 
Whether individually or taken together, these hints are not convincing. However, they are interesting enough to call for further, hopefully conclusive, investigation.

From the measured rate at which the Standard-Model (SM) $Z$ boson decays into invisible particles, we know that only three distinct flavors of neutrinos couple to the $Z$. Thus, if there are, say, $3 + n$ neutrino mass eigenstates, then $n$ orthogonal linear combinations of them are neutrino flavors that do not couple to the $Z$. Given the structure of the SM, neutrino flavors that do not couple to the $Z$ do not couple to the $W$ either. 
Such neutrino flavors, which do not experience any of the known forces of nature except gravity, are referred to as ``sterile'' neutrinos. As we see, the hints that there are more than three neutrino mass eigenstates are --- at the same time --- hints that nature contains sterile neutrinos. We note that although sterile neutrinos do not couple to the SM $W$ or $Z$ bosons, they may well couple to some as-yet-undiscovered non-SM particles. The latter particles could perhaps be found at the LHC or elsewhere.

The first hint that there are sterile neutrinos came from the LSND experiment, which reported evidence of a {\em rapid} $\nmbar \ra \nebar$ oscillation. \cite{refC} Assuming that this oscillation is sensitive to just one relatively large squared-mass splitting $\Dm2$, its probability, $P(\nmbar \ra \nebar)$, is given by
\be
P(\nmbar \ra \nebar) = \sin ^2 2\theta \sin^2 [1.27 \Dm2 (\mathrm{eV}^2) \frac{L(\mathrm{m})}{E(\mathrm{MeV})} ] ~~.
\label{eq1}
\ee
Here, $\sin^2 2\theta$ is a mixing parameter that satisfies $0 \le \sin^2 2\theta \le 1, \; L$ is the distance travelled by the antineutrinos between birth and detection, and $E$ is their energy. The antineutrinos studied by LSND came from $\mu^+$ decay at rest, so they had $E \sim 30$ MeV. They travelled $L \sim 30$~m. Thus, the LSND oscillation, which was found to have $P(\nmbar \ra \nebar) \sim 0.3\%$, occurred when $L$(m) / $E$(MeV) $\sim 1$. From \Eq{1}, such an oscillation requires a splitting $|\Dm2 | \gsim 0.1$ eV$^2$. 
Such a splitting is significantly larger than the two splittings $|\Dm2_{32}|$ and $|\Dm2_{21}|$ between the mass eigenstates of the three-neutrino paradigm. Thus, if the LSND oscillation is real, then the neutrino squared-mass spectrum obviously contains at least four neutrino mass eigenstates. Hence, from the data on $Z$ boson decays, there must be at least one sterile neutrino.

A second, related, hint of sterile neutrinos comes from the MiniBooNE experiment. In MiniBooNE, $L$ and $E$ are both $\sim 17$ times larger than they were in LSND, but the ratio $L/E$, on which the oscillation probability depends ({\it cf.} \Eq{1}), is comparable in the two experiments. MiniBooNE has searched for both $\nu_e$ appearance in a $\nu_\mu$ beam (i.e., $\nu_\mu \ra \nu_e$ oscillation), and $\nebar$ appearance in a $\nmbar$ beam ($\nmbar \ra \nebar$ oscillation). Its results have evolved over the years as more data have been accumulated and analyses have been refined. 
As of this writing (June 30, 2012), the $\nu_\mu \ra \nu_e$ and $\nmbar \ra \nebar$ data look fairly similar. An excess of electron-like events above background is seen in both the neutrino and antineutrino beams. The data are found to be compatible with the hypothesis of $\nu_\mu \ra \nu_e$ and $\nmbar \ra \nebar$ oscillation with $\Dm2 \gsim 0.1$ eV$^2$. In addition, the data are qualitatively similar in appearance to the LSND $\nmbar \ra \nebar$ data. \cite{ref1}

A third hint of sterile neutrinos has come from reactor antineutrino experiments in which the detector is relatively close to the reactor --- between 10 m and 100 m. The theoretical prediction for the un-oscillated $\nebar$ flux from reactors has recently increased by $\sim 3\%$. \cite{ref2} When this new prediction is used, the reactor experiments with close detectors show a $\sim 6\%$ deficit in the measured $\nebar$ flux. 
This deficit suggests a disappearance of $6\%$ of the reactor antineutrinos, all of which are born as $\nebar$, through oscillation into other flavors. Given the (10 - 100) m distance $L$ to the detector, the $\sim 3$ MeV typical energy $E$ of a reactor antineutrino, and the characteristic $\sin^2  [1.27  \Dm2 (\mathrm{eV}^2) L(\mathrm{m}) / E(\mathrm{MeV}) ] $ dependence of oscillation on $L/E$ illustrated by \Eq{1}, we see that the apparent disappearance of flux points to a splitting $\Dm2 \gsim 0.1$ eV$^2$, like the splitting suggested by LSND and MiniBooNE. Oscillation based on a four-neutrino spectrum in which one $\Dm2$ is much larger than 1 eV$^2$ describes the reactor data well.

An additional hint of sterile neutrinos has come from observation of the $\nu_e$ fluxes produced by intense $^{51}$Cr and $^{37}$Ar sources that are placed inside gallium detectors. The rate of events induced by these $\nu_e$ fluxes is found to be $\sim 15\%$ less than predicted. \cite{ref3} Perhaps this is due to a disappearance of $\nu_e$ flux caused by oscillation into other flavors. If so, the oscillation is occurring at $L$(m) / $E$ (MeV) values of order unity. Thus, once again we have a hint of a $\Dm2 \gsim 0.1$ eV$^2$ --- a large splitting that calls for a fourth neutrino mass eigenstate.

Besides these hints from terrestrial experiments, there is a hint of at least one sterile neutrino coming from cosmology. Big Bang Nucleosynthesis (BBN) and Cosmic Microwave Background (CMB) anisotropies count the effective number of relativistic degrees of freedom, $N_{\mathrm{eff}}$, at early times when light neutrinos would have been at least somewhat relativistic. An active neutrino (i.e., one that couples to the $W$ and $Z$), and a sterile neutrino that mixes with the active ones as required by the terrestrial hints, would both very likely have thermalized in the early universe. 
Then each of them would have made a contribution of unity to $N_{\mathrm{eff}}$. Now, BBN and CMB observations show that $N_{\mathrm{eff}}$ may be closer to 4 than to 3. \cite{ref4} This, in turn, suggests that, in addition to the mass eigenstates $\nu_1,\; \nu_2$, and $\nu_3$ of the standard three-neutrino paradigm, there may be a fourth light mass eigenstate, and consequently a sterile neutrino. (To be sure, it could be that there is an extra relativistic degree of freedom, but it is not a neutrino.) More precise information on $N_{\mathrm{eff}}$ will come from the Planck satellite in approximately one-half year.

The extent to which one can simultaneously fit all the oscillation data from terrestrial experiments, including experiments that do not show any evidence for extra neutrino states, with four- and five-neutrino spectra has been explored. In general, the four-neutrino spectrum is taken to have the mass eigenstates $\nu_{1,2,3}$ of the standard three-neutrino paradigm adjacent to each other, and one additional mass eigenstate $\nu_4$ some distance above or below all three of them (``$3+1$'' spectrum). Similarly, the five-neutrino spectrum is taken to have $\nu_{1,2,3}$ adjacent to each other, and both of two additional mass eigenstates, $\nu_4$ and $\nu_5$, above or below all three of them  (``$3+2$'' spectrum). 
All of the terrestrial experiments that show evidence of large squared-mass splittings, hence extra neutrino mass eigenstates, are ``short-baseline'' (small $L/E$) experiments. At the values of $L/E$ encountered in these experiments, the splittings $\Dm2_{21}$ and $\Dm2_{32}$ are too small to be visible, so $\nu_3,\; \nu_2$, and $\nu_1$ may be taken to be degenerate. Then the $3+1$ spectrum is sensitive to only one splitting, $\Dm2_{41}$, and the $3+2$ spectrum to only two independent splittings, $\Dm2_{41}$ and $\Dm2_{51}$.

Neither a $3+1$ nor a $3+2$ spectrum can simultaneously provide a good fit to all the oscillation data that exist as of this writing. The major difficulty is a tension between the $\optbar{\nu_\mu} \ra \optbar{\nu_e}$ appearance signals of LSND and MiniBooNE on the one hand, and the upper bounds on any $\nu_\mu$ and $\nebar$ disappearance signals in other experiments on the other hand. \cite{ref5} This tension reflects the fact that in both $3+1$ and $3+2$ models, appearance and disappearance probabilities are related. Perhaps the physics underlying short baseline oscillation is more complicated than that in the $3+1$ and $3+2$ models.

\section{The Future}

At the end of 2011, Fermilab created a Short-Baseline Neutrino Focus Group to consider how the tensions with the three-neutrino framework might be resolved. This Group (to which the present writer belongs) believes that these tensions are intriguing, and persistent enough to warrant definitive investigation. The Group has recommended a short-baseline plan for Fermilab that would include a new experiment to search for $\nu_\mu \ra \nu_e$ and/or $\nu_e \ra \nu_\mu$ transitions. If there should be a sterile neutrino discovery, the plan would also include a further experiment or experiments that could explore as many different flavor transitions as practical over the relevant $L/E$ range. \cite{ref6}

The community has put forward  a number of creative ideas for possible future experiments that could address the hints of sterile neutrinos.  Here, we would just like to illustrate the diversity of these ideas.

One idea is to study the coherent scattering of a neutrino from an entire nucleus. Such scattering would proceed via $Z$-boson  exchange. The $Z$ boson couples in a universal and flavor-preserving manner to the three active neutrino flavors, $\nu_e,\; \nu_\mu$, and $\nu_\tau$. However, as already discussed, the $Z$ boson does not couple to sterile neutrinos. Consequently, if an active neutrino, such as a $\nu_\mu$, oscillates into a sterile neutrino, the coherent scattering event rate will oscillate along with it. 
Such an oscillation would be quite striking. All the existing terrestrial hints of sterile neutrinos are kinematical; they are hints of a large splitting $\Dm2 \sim 1$ eV$^2$, and consequently of a {\em heavy} neutrino, with mass $\sim 1$ eV. In contrast, the observation of oscillation of the $Z$-exchange-induced coherent neutrino scattering event rate would be direct evidence of the existence of {\em sterile} neutrino flavors.

One specific suggestion for how coherent neutrino-nucleus scattering might be studied proposes to use the mono-energetic electron neutrinos from an $^{37}$Ar source, and to detect the very-low-energy nuclear recoils (with kinetic energy $\sim$ Few $\times 10$ eV) produced by the coherent scattering using cryogenic bolometers. \cite{ref7}

Another specific suggestion preposes to use a pion- and muon-decay-at-rest (DAR) neutrino source. \cite{ref8} Coherent scattering of the neutrinos from this source would produce nuclear recoils with kinetic energy $\sim 10$ keV, and detection via Dark-Matter-search-inspired detectors is considered.

A number of quite different kinematical studies that would look for oscillations at high $\Dm2$ have been proposed. One idea is to position a DAR source next to a very large liquid scintillator detector such as LENA or NO$\nu$A. \cite{ref9} For a neutrino with energy $E \sim 30$ MeV from a DAR source, a $\Dm2$ of 1 eV$^2$ would lead to an oscillation whose first maximum is at $\sim 40$ m, which is within the length of one of these large detectors. The oscillation could then be studied as a function of distance within the detector.

There have been several proposals for accelerator-based experiments that would compare event rates in a near and a far detector. This is a good way to deal with flux uncertainties, as long as the splitting $\Dm2$ is not so large that the neutrinos have already oscillated before reaching the near detector. It has been proposed to move the ICARUS detector from Gran Sasso to CERN to act as a far detector and to build a second detector to act as a near one. \cite{ref10} It has also been proposed to build an experiment in the Fermilab booster neutrino beam using the MicroBooNE liquid argon detector as the near detector, and a new, large liquid argon detector as the far one . \cite{ref10}

A particularly ambitious proposal is the idea of probing the existence of  high-$\Dm2$ oscillations using neutrinos from a very low energy neutrino factory. \cite{ref11} This factory would store positive muons with energies of (2-3) GeV in a storage ring, and look for the oscillation $\nu_e \ra \nu_\mu$ of electron neutrinos produced in the muon decays $\mu^+ \ra e^+ + \nu_e + \nmbar$. Assuming CPT invariance, the probability of the oscillation $\nu_e \ra \nu_\mu$ is the same as that of the oscillation $\nmbar \ra \nebar$ reported by LSND. The projected ability of the very low energy neutrino factory to confirm or exclude oscillations that are driven by a splitting $\Dm2 > 0.1$ eV$^2$, and are at the level suggested by LSND and MiniBooNE, is impressive.

In summary, there are interesting tensions with the three-neutrino paradigm. Hopefully, we will be able to determine what lies behind them in the not too distant future.

\section*{Acknowledgments}

We thank Alan Bross, Patrick Huber, Georgia Karagiorgi, and Joachim Kopp for helpful inputs, and the organizers of Moriond for many things. We appreciate partial support from the European Union FP7 ITN INVISIBLES (Marie Curie Actions, PITN- GA-2011- 289442).

\end{document}